\documentclass[10pt]{iopart} 
\usepackage[ansinew]{inputenc}
\usepackage[T1]{fontenc} 
\usepackage[english]{babel}
\usepackage{amsfonts} 

\usepackage{graphicx} 
\usepackage{epsfig} 
\usepackage{amssymb}
\newcommand{\Id}{e}
\begin{document}
\title{Explicit expressions for the topological defects of spinor 
Bose-Einstein condensates}
\author{H Mäkelä}  
\address{Department of Physics,
University of Turku,
FIN-20014 Turun yliopisto,
Finland}

\begin{abstract}
 In this paper we first derive a general method which enables one to create  
  expressions for vortices and monopoles.   
By using this method we construct several order-parameters 
describing the vortices and monopoles of Bose-Einstein condensates with 
hyperfine spin $F=1$ and $F=2$. 
We concentrate on defects 
which are topologically stable in the absence of an external magnetic 
field. In particular we show that in a ferromagnetic condensate there 
can be a vortex which does not produce any superfluid flow. 
We also point out that the order-parameter space of the cyclic phase of $F=2$ 
condensate consists of two disconnected sets. 
Finally we examine 
the effect of an external magnetic field on the vortices of a ferromagnetic $F=1$
 condensate  and discuss the experimental preparation of a vortex in this system.

\end{abstract}
\section{Introduction}
During the last ten years 
Bose-Einstein condensates (BECs) of alkali atoms  
have turned out to be an excellent system  
to create and observe several interesting phenomena, such as 
topological defects \cite{Fetter01,Kasamatsu05}. 
The best-known topological defect is a vortex, 
which in a typical single component 
BEC appears as a long-lived line-like singularity in the particle
density. 
In a non-rotating trap, a   
vortex state cannot be energetically the ground state of the system, 
but its decay is prevented 
by topological reasons. The continuous deformations of the
order-parameter which are needed in order to reach the ground state  
require more energy than what is available from e.g. thermal
excitations.   
In the presence of dissipation the vortex 
can move to the boundary of the condensate and vanish, but even then 
it is stable as long as it stays in the condensate.  
The creation of spinor BECs has made it  
possible to have more complicated vortices and other topological 
structures than what are allowed by a single component condensate 
\cite{Kasamatsu05,Makela03,Ho98}. 
By spinor condensates we mean BECs which have all spin components 
trapped simultaneously and where spin dynamics between different spin components 
is possible. For these reasons spinor condensates allow for richer topological structures 
than single component condensates.  
In experiments spinor condensates are realized by using 
an optical trap to trap the condensed atoms. 
If a BEC is in a magnetic trap, 
only particles which 
are in a low-field seeking state with respect to the quantization 
axis determined by the local magnetic field remain trapped.   
When an optical trap is used 
there can still be magnetic fields present. They are not needed 
to trap atoms, but e.g. to diminish the effect of stray magnetic
fields \cite{Stenger98,Schmaljohann04,Kuwamoto04,Chang04}.

The existence 
of the topological defects is based on the fact that 
a BEC can be described by an order-parameter $\psi$. 
The order-parameter is a map from some
region of physical space into the order-parameter space $M$. 
By examining the properties of the order-parameter 
space one can see what kind of, if any, topological defects are 
possible. This examination 
can be carried out with the help of the homotopy groups 
of the order-parameter space. 

The use of the theory of homotopy groups to characterize the
 topological defects of physical systems was first used 
during the late fifties \cite{Finkelstein59}, but 
 became widely known only in the seventies; see  e.g. \cite{Mermin79}. 
Since then it has been   
successfully applied in several fields of physics, such as
 condensed matter physics, particle physics and cosmology
 \cite{Mermin79,Trebin82,Rajaraman82,Vollhardt90,Vilenkin94,Volovik03}. 
Homotopy groups classify maps which can be continuously 
deformed into one another. Homotopy is a mathematical notion giving an exact 
meaning for this kind of deformation. 
Homotopy groups have turned out to be an effective way of 
 characterizing and classifying 
topological defects. This classification can be achieved if the  
order-parameter space $M$ is identified with a quotient 
space $G/H$, where $G$ is a group that acts 
transitively on the order-parameter space and $H$ is a suitably 
chosen subgroup of $G$.

If $G/H$ is known, information on the topological 
defects can be obtained by calculating the 
homotopy groups $\pi_n(G/H)$, $n=0,1,2,3$. From these $\pi_0(G/H)$ 
characterizes domain walls, $\pi_1(G/H)$ vortices and one-dimensional
non-singular defects, $\pi_2(G/H)$ 
monopoles and two-dimensional non-singular defects, and 
$\pi_3(G/H)$ three-dimensional non-singular defects. 
Domain walls, vortices and 
monopoles are defects where the order-parameter has to  vanish 
at some point of the physical space.   
The non-singular defects are defects where the 
order-parameter is nonzero everywhere, and the topological stability 
is imposed by the boundary conditions. 
The elemets of the homotopy 
group label the order-parameters in such a way 
that those labelled by the same group element 
can be continuously converted into one another, 
whereas if the configurations are labelled by different 
group elements, this is not possible. If $\pi_n(G/H)=\{\Id\},$ 
 i.e. the $n$th  homotopy group is a one element group, 
no topologically stable defects characterized by 
 $\pi_n(G/H)$ are possible.

This paper is organized as follows. 
In section 2 we derive a systematic way to find expressions 
for vortices and monopoles. This method is based on the 
properties of the relative homotopy groups, and it has not been presented 
before. 
In section 3 we review the properties of spinor condensates 
and their ground states. As a new result we show that the order-parameter space 
of the cyclic phase of $F=2$ condensate consists of two sets which are disconnected. This is in 
contrast to the order-parameter spaces of other ground state phases, which consist of one connected set. 
In section 4 the method 
derived in section 2 is applied in the 
context of spinor BECs to create order-parameters describing 
vortices and monopoles. Most of the expressions for defects 
have not been presented before. 
These are then used 
to find the minimum energy states of the defects and 
to study the superfluid velocity and angular momentum induced by them. 
In particular we show that in a ferromagnetic condensate the 
presence of a vortex does not have to lead to   
superfluid flow.
In section 5 we show that the vortex of a ferromagnetic $F=1$ 
condensate derived in section 4 can exist also if 
the conservation of magnetization is taken into attention.
Additionally we show that in a ferromagnetic $F=1$ condensate the conservation of magnetization 
may lead to stabilization of defects which are not stable if the magnetization 
is allowed to vary freely.    
In section 6 we propose a way to create 
vortices in a ferromagnetic condensate. 
These vortices are stable in the absence of an external magnetic field. 
This way is based on the use of a topological vortex creation 
method used before to create defects which are 
stable in the presence of an external field. 
Finally in section 7 we summarize the results 
of the paper and in appendix A we give the spin and rotation 
matrices needed in the calculation of defects.

\section{Finding expressions for vortices and monopoles}
\subsection{Introduction}
In this section we develop a method for finding explicit expressions 
for vortices and monopoles. The content of this section is quite
mathematical, and those interested only in the applications may jump
to the next section.  
We briefly introduce some of the necessary concepts and fix 
the notation.  
To save space 
we do not define all mathematial conceps needed; an interested reader
 can find the definitions  
in any standard book on homotopy theory, such as \cite{Hatcher02}. 
We also omit all proofs 
here; those can be found in \cite{Mermin79,Steenrod51}.   
From now on all maps are assumed to be continuous. 
In the following $G$ is a Lie group and $H$ is a closed subgroup of
$G$. 
The subgroup $H$ can be written as  
$H=H_0\cup H_1\cup H_2\cup\cdots$, where 
$H_i$'s are the path components of $H$. 
The path component containing the 
identity element $\Id$ is denoted by $H_0$. It is a normal subgroup of $H$ 
and thus the quotient space $H/H_0$ is a group. 
The path components of $H$ are
cosets of $H$ in $H_0$, i.e.  $H_k=h_k H_0$ for some
 $h_k\in H$. We define the $n$-dimensional disc by 
$D^n=\{\mathbf{x}\in\mathbb{R}^n\, |\, |\mathbf{x}|\leq 1\,\}$ and the 
$n$-sphere by  
 $S^{n} =\{\mathbf{x}\in\mathbb{R}^{n+1}\, |\, |\mathbf{x}|= 1\,\}$. 
We denote the $n$th relative homotopy by $\pi_n(G,H,\Id)$ and 
an element of this group by $[f]$, where $f:(D^n,S^{n-1},s_0)\rightarrow (G,H,\Id)$ is a map.
Here the notation means that $D^n$ is taken to $G$, $S^{n-1}$ to $H$, and 
$s_0\in S^{n-1}$ to $\Id$ by $f$.  By $[f]$ we denote the equivalence class determined  by $f$. 
It consists of all maps which are homotopic via maps of this type. 
If $H=\Id$, we write $\pi_n(G,\Id)\equiv\pi_n(G,\Id,\Id)$. 
If $G$ is path connected,  $\pi_n(G,g_1)$ and $\pi_n(G,g_2)$ are isomorphic 
for all $g_1,g_2\in G$, and we use the notation $\pi_n(G)$ to denote 
any $\pi_n(G,g_1)$. 
There is an exact sequence of homomorphisms between relative homotopy groups. 
This sequence reads 
\begin{equation}\label{Exact2}
\put(-70,0){\vector(1,0){16}}
\put(-70,5){$\alpha_n$}
\put(-50,-2){$\pi_n(G,\Id)$}
\put(-10,0){\vector(1,0){16}}
\put(-10,5){$\beta_n$}
\put(10,-2){$\pi_n(G,H,\Id)$}
\put(28,-6){\vector(0,-1){16}}
\put(10,-35){$\pi_n(G/H,H)$}
\put(35,-15){$p_*$}
\put(62,0){\vector(1,0){16}}
\put(62,5){$\gamma_n$}
\put(82,-2){$\pi_{n-1}(H,\Id)$}
\put(132,0){\vector(1,0){16}}
\put(132,5){$\alpha_{n-1}$}
\put(152,-2){$\pi_{n-1}(G,\Id)$} 
\put(202,0){\vector(1,0){16}}
\put(202,5){$\beta_{n-1}$}
\put(222,-2){$\cdots$}
\end{equation}
 We have included in the sequence $p_*$, which is 
an isomorphism determined by the map
 $p:G\rightarrow G/H, g\mapsto gH$. If $[f]\in \pi_n(G,H,\Id)$ then  
$p_*([f])= [pf]\in\pi_n(G/H,H)$.

\subsection{Physical applications}
Next we consider how  previous results can be applied in physical
systems. Now  $G$ is a group that acts transitively on 
the order-parameter space $M$. The action 
of $g\in G$ on $x\in M$ is denoted by $g\cdot x$. 
We now choose 
an arbitrary element $x_{ref}\in M$, called the reference
order-parameter, and define 
$H=\{g\in G\,|\,g\cdot x_{ref}=x_{ref}\}$. This is the isotropy group 
and it is a closed subgroup of $G$. 
The order-parameter space $M$ can then be identified 
with $G/H$. The correspondence between the elements of $M$ and 
$G/H$ is $x\Leftrightarrow gH, g\cdot x_{ref}=x$. 
If $A$ is a topological space and $f:A\rightarrow G$ is a map,  
then the map $pf:A\rightarrow G/H$ gives 
a map from $A$ to the order-parameter space, 
which is now represented by $G/H$. If the  
order-parameter space is represented by $M$ the corresponding map 
from $A$ to $M$ is given by 
$c:A\rightarrow M$ such that $c(a)
=f(a)\cdot x_{ref}$ for all $a\in A$. When discussing the physical 
applications we assume that
$\pi_2(G,\Id)=\pi_1(G,\Id)=\pi_0(G,\Id)=\{\Id\}$, when  
$\gamma_1$ and $\gamma_2$ become isomorphisms. These conditions  
hold for $\mathbb{R}$ and $SU(2)$, which are the groups 
used in this paper.

\subsubsection{Monopoles}

Because $\pi_1(H,\Id)=\pi_1(H_0,\Id)$, we  
see that $\pi_2(G,H,\Id)$ and $\pi_1(H_0,\Id)$ are isomorphic via
$\gamma_2$. The map $\gamma_2p_*^{-1}$ 
gives an isomorphism between $\pi_2(G/H,H)$ and $\pi_1(H_0,\Id)$. 
This isomorphism can be used in the calculation of $\pi_2(G/H,H)$, 
since usually it is quite easy to see what $\pi_1(H_0,\Id)$ is 
\cite{Makela03,Mermin79,Trebin82}. 
Assume that $[f]\in \pi_2(G,H,\Id)$  and that $f\big|_{S^{1}}$  is 
the  restriction of $f$  to $S^{1}$, the boundary of $D^2$.  
Then $[f\big|_{S^{1}}]\in \pi_{1}(H_0,\Id)$. 
The isomorphism 
$\gamma_2$ is given by the map $[f]\mapsto[f\big|_{S^{1}}]$.  
We use 
$(\theta,\varphi)$ as the coordinates of $D^2$, that is,  $\theta$ is
the distance from the center of the disc and $\varphi$ is the azimuthal
angle. We take the radius of $D^2$ to be $\pi$. Let
$g:[0,2\pi]\rightarrow H_0$ be a map for which $g(0)=g(2\pi)=\Id$ and let 
$\tilde{g}:D^2\rightarrow G$ be such that  
 $\tilde{g}(\theta=\pi,\varphi)=g(\varphi)$. 
If we also define $s_0=(\pi,0)$,    
then $[\tilde{g}]\in\pi_2(G,H_0,\Id), [p\tilde{g}]\in\pi_2(G/H,H)$ and 
$[\tilde{g}\big |_{S^1}]=[\tilde{g}\big|_{\theta=\pi}]=[g]
\in\pi_1(H_0,\Id)$. As explained above, $[pg]$ and $[g]$ 
are mapped into each other by isomorphism $\gamma_2p_*^{-1}$. 

Next we construct a map from physical space into the order-parameter 
space $G/H$ which describes a monopole with some given winding
number. 
Now we use spherical coordinates $(r,\theta,\varphi)$ as the 
coordinates of the physical space $\mathbb{R}^3$ and assume that the 
monopole is located at the origin. The monopole we construct is
independent of the $r$-coordinate. This assumption is not
necessary, but 
to avoid further complication we use it here. 
Let 
$A =\mathbb{R}^3\setminus \{\mathbf{0}\}$. We define $f:A\rightarrow
G/H$ by $f(r,\theta,\varphi)=\tilde{g}(\theta,\varphi)H$.   
Then for each fixed $r>0$  $[f]\in\pi_2(G/H,H)$ and $[f]$ is  
the unique inverse image of $[g]
\in\pi_1(H_0,\Id)$ in the map $\gamma_2p_*^{-1}$. 
These elements have the same winding number.  
Thus if one wants a map $f:A\rightarrow G/H$, 
which describes a monopole with a given winding
number, he has to find $\tilde{g}:D^2\rightarrow G$ such that  
$[\tilde{g}\big|_{\theta=\pi}]$ is an element of $\pi_1(H_0,\Id)$ with the
wanted winding number. Then $f$ is obtained by defining
$f(r,\theta,\varphi)=\tilde{g}(\theta,\varphi)H$.  
The corresponding order-parameter $x:A\rightarrow M$ 
is defined by 
$x(r,\theta,\varphi)=\tilde{g}(\theta,\varphi)\cdot x_{ref}$.

\subsubsection{Vortices} 

The group  structure of $\pi_0(H,\Id)$ 
has to be defined a little differently than that of other relative 
homotopy groups \cite{Mermin79}. We  
define $\pi_0(H,\Id)\equiv H/H_0=\{H_0, H_1, H_2,\ldots \}$ and 
$D^1=[0,2\pi]$. 
 Let $g:D^1\rightarrow G$ be such that
$g(0)=\Id$ and  $g(2\pi)\in H_m$ for some $m\in \{0,1,2,\ldots\}$. 
If we choose  $s_0=0$ then $[g]\in 
\pi_1(G,H,\Id)$. The isomorphism $\gamma_1:\pi_1(G,H,\Id)
\rightarrow \pi_0(H,\Id)$ is given by $[g]\mapsto [g(2\pi)]$, 
where $[g(2\pi)]\equiv H_m$.

We use cylidrical 
coordinates $(r,z,\varphi)$ as the coordinates of physical 
space and assume that the vortex 
is located on the $z$-axis. We define $A=\mathbb{R}^3\setminus
\mathbb{R}\mathbf{e}_z$, where $\mathbb{R}\mathbf{e}_z$ denotes the $z$-axis. 
We define $f:A\rightarrow G$ such that $f(r,z,0)=\Id$ and $f(r,z,2\pi)\in
H_m$. Then  
for each fixed $(r,z)$ $[f]\in\pi_1(G,H,\Id),\,\, [pf]\in\pi_1(G/H,H)$ 
and the image of $[f]$ in the isomorphism 
$\gamma_1$ is $[f(r,z,2\pi)]=H_m\in\pi_0(H,\Id)$. 
Thus $f$ gives a vortex with the winding number represented by  
$H_m\in\pi_0(H,\Id)$. The corresponding order-parameter 
$x:A\rightarrow M$ is defined
by $x(r,z,\varphi)=f(r,z,\varphi)\cdot x_{ref}$. For a vortex 
constructed this way $x(r,z,0)=x_{ref}$ for all $r>0,z\in\mathbb{R}$. This 
requirement can be relaxed, but for our purposes that  
is not necessary.   

\section{Spinor Bose-Einstein condensates}  
 
In this section we review the ground-state order-parameter spaces 
of spinor BECs. As a new result we show that the order-parameter space of the 
cyclic phase consists of two disconnected sets. A spinor in one set cannot 
be continuously converted into a spinor in the other set while staying 
in the order-parameter space all the time. Only the order-parameter space of the 
cyclic phase has this structure, since 
in other ground states the order-parameter space is connected.

A spinor condensate of atoms with hyperfine spin equal to $F$, 
$F=1,2,\ldots$, is described in 
the mean field theory by the order-parameter $\psi$, 
which can be written in
the form  
$\psi(\mathbf{r})=\sqrt{n(\mathbf{r})}\xi(\mathbf{r}),$ 
where $n(\mathbf{r})$ is the particle density, 
$\xi(\mathbf{r})$ is the transpose of the complex vector
 $(\xi_F(\mathbf{r}),\xi_{F-1}(\mathbf{r}),
\ldots ,\xi_{-F}(\mathbf{r}))$ 
and $\xi(\mathbf{r})^\dag\xi(\mathbf{r})=1$ \cite{Ho98,Ohmi98}. 
In the rest of the paper we call $\xi$ the spinor. In some 
publications the term spinor refers to $\psi$, so one must be careful 
with the terminology. When we determine the order-parameter space, 
we set the density equal to one, so  
 the order-parameter is just the spinor $\xi$\footnote{
In principle 
 the order-parameter space should be written as 
$M'=(\mathbb{R}_+\times M)\cup \{0\}$, where $\mathbb{R}_+$ is the set of
real numbers larger than zero giving the possible values of the square root 
of the density, and $M$ gives the order-parameter space related to  
the density-independent part of 
the order-parameter. Now $M$ consists of 
the possible values of the spinor, while in the case of a single
component condensate $M=S^1$, which characterizes the possible values 
of the phase of the order-parameter. 
 The point $\{0\}$ denotes the case where the
density is zero.  
One sees that if $M'$ is the order-parameter space there are no
topologically stable defects. Any order-parameter can be converted
into any other order-parameter via deformations which  reduce
 the density to zero in an appropiate region of the physical space. From
 a physical point of view 
this is unlikely to happen, since reducing density to zero is
not energetically favorable. In principle the necessary energy could 
come for example from thermal excitations, but in practise this is
unlikely to happen. 
Thus one can ignore the zero of density. Then the order-parameter 
space becomes $\mathbb{R}_+\times M$. For this  
$\pi_n(\mathbb{R}_+\times
M)=\pi_n(\mathbb{R}_+)\times\pi_n(M)=\pi_n(M),$ since
 $\pi_n(\mathbb{R}_+)=\{\Id\}$. Thus, from the point 
of view of topological defects, it is enough to study the structure of
$M$ only.}. 
If the normalization of $\xi$  is the only restriction imposed on 
the order-parameter, the order-parameter space becomes $S^{4F+1}$. This 
space allows for no topological defects characterized by $\pi_n$ with 
$n=0,1,2,3$. This is because from the theory of homotopy groups it is 
known that  $\pi_n(S^{4F+1})=\{\Id\}$ for $n=0,1,2,3$ and 
$F=1,2,3,...$.
However, for example the order-parameter space of an $F=1$ condensate  
may be only a subset of $S^5$. This can be inferred from the energy
functional, which for an $F=1$ condensate in the absence of external
 magnetic field reads \cite{Ho98,Ohmi98}
\begin{equation}\label{EF1}
\fl
E[\psi]=\int d^3\mathbf{r}\Big\{\frac{\hbar^2}{2M}
\sum_{i=-1}^{+1}\nabla\psi_i^*(\mathbf{r})\cdot\nabla
\psi_i(\mathbf{r})+V(\mathbf{r})n(\mathbf{r})
+\frac{n(\mathbf{r})^2}{2}[
\alpha_1+\beta_1\langle\mathbf{F}\rangle^2]\Big\}.
\end{equation}
 Here $\mathbf{F}$ is the (hyperfine) spin operator, $\langle 
\mathbf{F}\rangle=
\xi^\dag(\mathbf{r}) \mathbf{F}\xi(\mathbf{r})$, $M$ is the mass
 of the atom, 
 $V$ is the external potential, $\alpha_1=\frac{4\pi\hbar^2}{M}\frac{a_0+2a_2}{3}, 
\beta_1=\frac{4\pi\hbar^2}{M}\frac{a_2-a_0}{3}$, 
and $a_F$ is the s-wave 
scattering length in the total spin $F$ channel. 

One sees that it is energetically favoured that either  
$|\langle \mathbf{F}\rangle|=0$ or $|\langle\mathbf{F}\rangle|=1$, 
corresponding to the cases $\beta_1>0$ and $\beta_1 <0$, respectively. 
When $\beta_1>0$ the system is said to be  antiferromagnetic, 
whereas if $\beta_1 <0$ the system is said to be ferromagnetic. 
It turns out that a group that acts transitively on the set 
of spinors fulfilling the condition $|\langle\mathbf{F}\rangle|=1$ is
 $SU(2)$. This can be shown using a similar reasoning as that shown below 
 in the context of the ground-state phases of $F=2$ condensates.  
$SU(2)$ acts via its irreducible three-dimensional representation, 
i.e. the spin rotations of an $F=1$-particle. 
For spinors with $|\langle\mathbf{F}\rangle|=0$ we choose 
$G=\mathbb{R}\times SU(2)$, where $\mathbb{R}$ gives 
the gauge transformations of the spinor as $\xi\mapsto 
e^{i\theta}\xi$, $\theta\in\mathbb{R}$.    
As before, $SU(2)$
describes spin rotations.

For an $F=2$ condensate the energy functional is \cite{Ciobanu00}  
\begin{equation}\label{EF2}
\fl
E[\psi]=\int d^3\mathbf{r}\Big\{\frac{\hbar^2}{2M}
\sum_{i=-2}^2\nabla\psi_i^*(\mathbf{r})\cdot\nabla
\psi_i(\mathbf{r})+V(\mathbf{r})n(\mathbf{r})
+\frac{n(\mathbf{r})^2}{2}(\alpha_2+\beta_2 \langle\mathbf{F}\rangle^2
+\gamma_2 |\Theta(\mathbf{r})|^2)\Big\}.
\end{equation}
Here $\alpha_2=\frac{1}{7}(4g_2 +3g_4), \beta_2=-\frac{1}{7}(g_2
-g_4)$, and  
$\gamma_2=\frac{1}{5}(g_0-g_4)-\frac{2}{7}(g_2 -g_4)$, where 
$g_i=\frac{4\pi\hbar^2 a_i}{M}$  and  
 $\Theta=2\xi_2\xi_{-2}-2\xi_1\xi_{-1}+\xi_0^2$. 
As in an $F=1$ system, the energy is invariant in position-independent 
spin rotations and gauge transformations.  
The possible ground states have been calculated in 
\cite{Ciobanu00,Ueda02}, and can be classified as follows. 
(i) If $\beta_2, \gamma_2>0$ the energy is minimized when 
$|\langle\mathbf{F}\rangle|=\Theta=0$. Spinors 
with these properties are called cyclic. (ii) When $\beta_2<0, \gamma_2>0$ 
the minimum is obtained by making $|\langle\mathbf{F}\rangle|=2,\Theta=0$, 
and the ground state is ferromagnetic. (iii) If $\beta_2>0, \gamma_2<0$ 
the minimum 
is achieved by maximizing $\Theta$, i.e. $|\Theta|=1$, and 
$|\langle\mathbf{F}\rangle|=0$.  
The ground state is called polar or antiferromagnetic.  
(iv) Finally, if $\alpha_2$ and $\beta_2$ are both negative, the 
 ground state is ferromagnetic for $4|\beta_2|>|\gamma_2|$, 
and polar otherwise. This is because $|\langle\mathbf{F}\rangle|$ and 
$|\Theta|$ cannot be maximized simultaneously.

Next we discuss briefly the structures of the order-parameter spaces 
of the ground states.   
Since the interaction energy is invariant 
in gauge transformations and spin rotations and $\beta_2,\gamma_2$ are arbitrary, also 
$|\Theta|$ and $\langle\mathbf{F}\rangle^2$ are invariant in gauge transformations and 
spin rotations. Furthermore, for every spinor $\xi$ there is always a spin rotation $R$ 
which rotates the spinor so that in the rotated state $R\xi$ the 
spin is parallel to the $z$-axis, i.e. 
$\langle\mathbf{F}\rangle=(R\xi)^\dag\mathbf{F}R\xi=
(R\xi)^\dag F_z R\xi \mathbf{e}_z=\langle F_z\rangle\mathbf{e}_z$. 
In the rotated state $\langle F_x\rangle=\langle F_y\rangle=0$.
Thus the spin-dependent terms in equation (\ref{EF2}) can be written 
as $\beta_2 \langle F_z\rangle^2+\gamma_2 |\Theta|^2$.
In the cyclic phase the energy is minimized when 
$\langle F_z\rangle= |\Theta|=0$.  Numerical 
calculations show that the solutions to these  
equations 
are (up to a gauge transformation and a rotation 
about the $z$-axis) $C0=\frac{1}{2}(1,0, \sqrt{2},0,-1)^T$, $C1=\frac{1}{\sqrt{3}}(1,0,0,\sqrt{2},0)^T$ 
and $C1'=\frac{1}{\sqrt{3}}(0,\sqrt{2},0,0,1)^T$ \cite{Sengstock04,Saito05}. 
All cyclic spinors can be obtained from these spinors by a 
gauge transformation and a spin rotation.
By exploiting the rotation matrix shown in the appendix one can show 
that $C1,C1'$ can be rotated into each other, while for $C0,C1$ and 
$C0,C1'$ this is not possible. This means 
that the order-parameter space of the cyclic phase 
consists of two disconnected sets, a fact that has not been pointed
out before. Formally this can is expressed 
as $\pi_0(G/H)=\mathbb{Z}_2$. The group  $G=\mathbb{R}\times SU(2)$ 
acts transitively on both disconnected sets. 
In the ground-state 
the system can consist of regions, some of which  
are in a $C0$ state and others are in a $C1$ 
state. These regions are separated by a domain wall. This kind of structure 
is not possible in other zero field ground-state 
phases of  $F=1$ or $F=2$ condensates. 

In the case of ferromagnetic ground states, the situation is simpler. 
Now the equations to be solved are 
$|\Theta|=\langle F_x\rangle=\langle F_y\rangle=0$ and $|\langle F_z\rangle|=2$. 
The only solutions (up to a phase) are $|F=2, m_F =2\rangle$  and $|F=2, m_F =-2\rangle$.  
These spinors can be rotated into each other, so the order-parameter space is now connected. 
By examining the spinors obtained by a spin rotation from 
the reference order-parameters one sees that 
in the case of a ferromagnetic condensate we can choose $G=SU(2)$ instead of $\mathbb{R}\times SU(2)$.  

Similar study for the polar phase shows that the order-parameter space  
of the polar phase is connected but larger 
than $\mathbb{R}\times SU(2)$. This complicates 
the study of polar defects, and they will not 
be discussed here \cite{Makela03}.

Before going to the details of defects we 
review the possible ground-state phases of some  
alkali atom condensates. 
 Experimental and theoretical results indicate that the 
$F=1$ spinor Bose-Einstein condensate of ${}^{87}$Rb is 
ferromagnetic \cite{Ho98,Schmaljohann04,Chang04,Klausen01}. 
On the other hand, the $F=2$ condensate of 
${}^{87}$Rb is probably  polar \cite{Schmaljohann04,Kuwamoto04,Klausen01}.
The ${}^{85}$Rb $F=2$ condensate seems to be polar \cite{Klausen01} 
and the ${}^{83}$Rb isotope with $F=2$ ferromagnetic \cite{Ciobanu00}.

${}^{23}$Na scattering lengths determined in 
\cite{Crubellier99} indicate that  ${}^{23}$Na $F=1$ 
is antiferromagnetic, as has been predicted in \cite{Ho98} and
 seen in experiments \cite{Stenger98}.
The ground state phase of ${}^{23}$Na $F=2$ spinor condensate  
 appears also to be antiferromagnetic \cite{Ciobanu00}. Experimental
 study of this condensate is difficult 
 because the $|F=2,m_F=0\rangle$ state  decays within 
milliseconds \cite{Gorlitz03}.

\section{Examples of defects in spinor condensates}
Next we present several examples of vortices and monopoles 
in spinor condensates.  
In these examples cylindrical coordinates $(r,z,\varphi)$ and spherical 
coordinates $(r,\theta,\varphi)$  are used when 
discussing vortices and monopoles, respectively. In vortices the
coordinate dependence is such that at $\varphi=0$ the reference spinor
is obtained. The 
vortex cores are assumed to be straight and located on the $z$-axis. Monopoles 
are located at the origin of the spherical coordinates. In what  
follows the minimum energy of a vortex with given winding 
number is also studied. This is calculated under the assumptions that the vortex 
stays fixed at the $z$-axis and the density $n$ is 
independent of $\varphi$. 
The ground state density can be assumed to be cylindrically symmetric   
if the external potential used to confine the condensate is 
cylindrically symmetric and the vortex does not split. 
The latter is not true in general, since usually if a condensate is trapped in a harmonic trap  
it is energetically favorable for a vortex with winding number larger than one to split 
into winding number one vortices. In this case the particle density cannot 
be cylindrically symmetric. The splitting can possibly be prevented by using a trap 
which is steeper than harmonic \cite{Lundh02} or by applying 
a repulsive potential in the vicinity of the rotation axis \cite{Pethick02}, p 245.

\subsection{Ferromagnetic condensates} 
Next we briefly examine the vortex configurations of ferromagnetic
condensates. Here we mean by ferromagnetic condensates 
systems whose ground state spinor is $|F,m_F=F\rangle$ or any other
spinor  
obtained from this by a spin rotation. A separate gauge transformation 
is now unnecessary, as a spin rotation alone is able to produce that. 
We choose $|F,m_F=F\rangle$ as our reference
order-parameter. The isotopy group is isomorphic 
with $\mathbb{Z}_{2F}$ and the order-parameter space                         
is $SU(2)/\mathbb{Z}_{2F}$; see \cite{Ho82}. The isotropy 
group for $F=1$ and $F=2$ has been calculated explicitly in
\cite{Makela03}. 
The winding numbers of topologically stable 
vortices range from $1$ to $2F-1$. If the winding 
number of a vortex is $m$, the winding number of the antivortex 
is $2F-m$. Especially, a vortex with winding number $F$ 
is also its antivortex.   
If $F=1$, the order-parameter space is 
$SO(3)$, since $SU(2)/\mathbb{Z}_{2}=SO(3)$ \cite{Ho98}. 
Because $\pi_2(SU(2)/\mathbb{Z}_{2F})=\{\Id\}$, in ferromagnetic condensates 
 monopoles are not topologically stable. 
Next we construct order-parameters describing vortices 
in ferromagnetic $F=1$ and $F=2$ condensates. The rotation matrices 
needed in the calculations can be found in appendix A. 

\subsubsection{Ferromagnetic $F=1$ condensate}

Now $H=\{\mathbb{I},-\mathbb{I}\}\in SU(2)$ and,  
as explained in section 2, a  vortex is 
given by a map $f:\mathbb{R}^3\setminus
\mathbb{R}\mathbf{e}_z\rightarrow SU(2)$ such that  
$f(r,z,0)=\mathbb{I}$ and  $f(r,z,2\pi)=-\mathbb{I}$ for every 
$z\in\mathbb{R},r>0$.  
From the $V$-matrix (\ref{V}) one sees that 
 this kind of map is obtained by choosing
 $\tau$ such that $\tau(r,z,0)=0$ and $\tau(r,z,2\pi)=2\pi$. The functions  
 $\alpha$ and $\beta$ can be arbitrary  functions of position.
The corresponding order-parameter is $\psi(\mathbf{r})=
\sqrt{n(\mathbf{r})}\xi(\mathbf{r})$, where 
$\xi(\mathbf{r})=f(\mathbf{r})\cdot\xi_{ref}
=V^{(1)}(\mathbf{r})\xi_{ref}$ and $V^{(1)}$ is given in equation (\ref{V1}).
We get 
\begin{equation}\label{F1}
\fl\psi(r,z,\varphi)=\sqrt{n(r,z,\varphi)}\left(
\begin{array}{cc}
&\left(\cos\frac{\tau}{2} - i\cos\beta\,\sin\frac{\tau}{2}\right)^2 \\
 &-\sqrt{2}e^{i\alpha}\sin\beta\,\sin\frac{\tau}{2}\,
 \left(i\cos\frac{\tau}{2}+\cos\beta\,\sin \frac{\tau}{2}\right) 
\\ &-e^{2i\alpha}\,\sin^2 \frac{\tau}{2}\sin^2\beta  
\end{array}\right).
\end{equation}  
From the way this was derived it follows that this is defined 
only for $r>0$. This order-parameter can, however, be extended to
whole $\mathbb{R}^3$ 
by requiring that $n(0,z,\varphi)=0$. This is needed 
 in order to keep the order-parameter well defined at $r=0$.

Next we try to find a spinor giving a vortex with the minimum energy. 
Now the superfluid velocity is 
$\mathbf{v}=-i\frac{\hbar}{M}\xi^\dag\nabla \xi$. 
When the vortex has the smallest possible energy 
the velocity in radial and $z$-directions can 
be assumed to vanish and thus the spinor is a function of $\varphi$ only.   
Additionally, the density is taken to be cylindrically symmetric.  
Using these assumptions the Euler-Lagrange equations obtained from the 
energy functional show that the vortex energy is minimized when 
$\alpha=constant,\beta=\frac{\pi}{2}$ and $\tau(r,z,\varphi)=\varphi$. 
Thus the vortex takes the form     
\begin{equation}\label{F1vortex}
\fl\psi(r,z,\varphi)=\sqrt{n(r,z)}\left(
\begin{array}{cc}
&\left(\cos\frac{\varphi}{2} - i\cos\beta\,\sin\frac{\varphi}{2}\right)^2 \\
 &-\sqrt{2}e^{i\alpha}\sin\beta\,\sin\frac{\varphi}{2}\,
 \left(i\cos\frac{\varphi}{2}+\cos\beta\,\sin \frac{\varphi}{2}\right) 
\\ &-e^{2i\alpha}\,\sin^2 \frac{\varphi}{2}\sin^2\beta  
\end{array}\right).
\end{equation} 
Here we have not set $\beta=\frac{\pi}{2}$, but have allowed it to have 
any value. This recalls a little the expression for the coreless vortex shown in \cite{Ho98}.
These are, however, different defects, since the vortex of equation 
(\ref{F1vortex}) cannot be converted into a non-vortex state by continuous 
deformations, whereas for the coreless vortex this can be done\footnote{The conservation 
of magnetization may change the situation. This is discussed in section 5.}. Another difference is that 
in the case of a coreless vortex the particle density does not vanish, while in the case 
of (\ref{F1vortex}) the density vanishes on the $z$-axis. 
The superfluid velocity for the vortex state (\ref{F1vortex}) is 
$\mathbf{v}=-\frac{\hbar}{M}\frac{\cos\beta}{r}\,\mathbf{e}_\varphi$ and  
if  $\beta =\frac{\pi}{2}$ it vanishes. 
Thus, in contrast to a single-component condensate, 
in a spinor condensate 
the existence of a vortex does not have to lead to a non-zero 
superfluid velocity. 
The same phenomenon  can be seen also in the 
orbital angular momentum. For the order-parameter (\ref{F1vortex}) it  
is $\mathbf{L}=-N\hbar \cos\beta\, \mathbf{e}_z$, which also vanishes  
if $\beta=\frac{\pi}{2}$. 
The condensate does not have 
to contain any angular momentum although there is a vortex 
in it. If the condensate is in a state with $\beta=0$, it contains  
one unit of angular momentum per particle and the system is 
similar to a single-component condensate with a vortex. 
In the presence  of 
dissipation the angular momentum
does not have to be conserved, and the vortex can evolve towards the
ground state where the angular momentum vanishes. 
The kinetic energy of (\ref{F1vortex}) is 
proportional to $3+\cos2\beta$, which decreases monotonically 
as $\beta$ increases from zero to $\frac{\pi}{2}$. Thus 
there is no energetic barrier which could render the vortex state 
with $\beta=0$ metastable against conversion into the ground state vortex.  

A vortex obtained from (\ref{F1vortex}) by setting $\beta=0$ has been
presented before in \cite{Ho98}. The general expression for the vortex
shown in equation 
(\ref{F1vortex}) or the behaviour of the angular momentum and
superfluid velocity have, however, not been discussed before. This
also holds true of the ferromagnetic $F=2$ vortices discussed next.

\subsubsection{Ferromagnetic $F=2$ condensate }

The isotropy group is 
$H=\{\mathbb{I},-i\sigma_z,(-i\sigma_z)^2,(-i\sigma_z)^3\}$, 
where $\sigma_z$ is the $z$-component of the Pauli matrices. 
A map $f_m:\mathbb{R}^3\setminus \mathbb{R}\mathbf{e}_z\rightarrow
SU(2)$ represents 
a vortex with winding number $m$ if $f_m(r,z,0)=\mathbb{I}$ and 
$f_m(r,z,2\pi)=(-i\sigma_z)^m$ for every $z\in\mathbb{R},r>0$.  
If we parametrize $SU(2)$ matrices as in (\ref{V}), these conditions are 
fulfilled if    
$\tau(r,z,0)=0$ and $\tau(r,z,2\pi)=m\pi$. Additionally,  
for $m=1,3$ the condition $\beta(r,z,2\pi)=2k\pi$ has to hold.  
Here $k$ is an arbitrary integer. For $m=2$ the function 
$\beta$ can be arbitrary. The order-parameter becomes  
\begin{eqnarray}\label{F2}
\fl\psi(r,z,\varphi)=\sqrt{n(r,z,\varphi)}
\left(
\begin{array}{c}
{\left( \cos  \frac{\tau }{2}-i\cos\beta\sin\frac{\tau }{2}\right)}^4\\
 2 e^{i  \alpha } \sin  \frac{\tau}{2}  \,\sin  \beta
 {\left( i  \cos\frac{\tau}{2}  + \cos\beta\sin\frac{\tau}{2}  
\right) }^3 \\
 {\sqrt{\frac{3}{8}}} e^{2i\alpha }  \sin^2  \beta\,
 {\left( \cos  \beta-\cos\tau\cos\beta + i\sin\tau\right) }^2 \\
 2 e^{3 i \alpha } \sin^3\frac{\tau}{2}\, \sin^3  \beta 
\left( i  \cos \frac{\tau}{2}  + \sin \frac{\tau}{2}\cos\beta
\right)  \\
e^{4 i\alpha } \sin^4  \frac{\tau}{2} \sin^4  \beta 
\end{array}\right).
\end{eqnarray} 
The minimum energy 
of a vortex with winding number $m$ is obtained when    
$\tau(r,z,\varphi)=\frac{m\varphi}{2}$ and $\xi=\xi(\varphi)$. 
If $m=2$ the vortex minimizing the energy 
is obtained from equation (\ref{F2}) by replacing 
$\tau$ with $\varphi$ and setting $\beta=\frac{\pi}{2}$. In this state,   
 the superfluid velocity and angular momentum vanish. We show that 
this is a general 
property of a winding number $F$ vortex in a ferromagnetic 
condensate with hyperfine spin $F$. This vortex is represented by the
element $-\mathbb{I}\in H$ and a vortex can be written as  
$\psi(r,z,\varphi)=\sqrt{n(r,z)}\xi(\varphi)$, where 
$\xi(\varphi)=e^{-i\varphi\, \mathbf{n}\cdot \mathbf{F}}\xi_{ref}$,  
$\mathbf{n}=(\cos\alpha\sin\beta,\sin\alpha\sin\beta,\cos\beta)$ 
is a constant vector and $\xi_{ref}=|F,m_F=F\rangle$. 
If $\beta=\frac{\pi}{2}$ we get $\mathbf{n}=(\cos\alpha,\sin\alpha,0)$ 
and $v_\varphi,L_z\sim \xi^\dag \frac{\partial}{\partial \varphi}\xi
=-i\xi_{ref}e^{i\varphi\, \mathbf{n}\cdot \mathbf{F}}\mathbf{n}\cdot \mathbf{F}
e^{-i\varphi\, \mathbf{n}\cdot \mathbf{F}}
\xi_{ref}=-i\xi_{ref}\mathbf{n}\cdot \mathbf{F}
\xi_{ref}=0$.  

If $m=1,3$ the function  $\beta$ that minimizes the energy of a vortex is 
more difficult to find. The obvious choice 
$\beta\equiv 0$ is not the correct one,  
since for example $\beta(\varphi)=\frac{8\pi +\varphi}{5}$  
produces a smaller energy. 
Finding the ground states of vortices with $m=1,3$ will be left 
for future publications, and will not be discussed here in more detail. 
If $\beta=0$ the vortex is simply   
$\psi(r,z,\varphi)=\sqrt{n(r,z)} 
e^{im\varphi}|F=2,m_F=2\rangle, m=1,3$.

\subsection{Antiferromagnetic $F=1$ condensate}
It is advantageous to use the $U$ matrix of equation (\ref{U})  
and corresponding representation matrices 
 when discussing the defects of 
an antiferromagnetic $F=1$ condensate.
The reference order-parameter is chosen to be $|F=1,m_F=0\rangle$ and 
 the isotropy group is $H=\{\big(
 m2\pi,U(\varphi,0,0)\big),
\big((m+\frac{1}{2})2\pi,g U(\varphi,0,0)\big)\,|\,\varphi \in
[0,4\pi],
 m\in \mathbb{Z}\}
 $, where we have defined $g=U(0,\pi,0)$.
 The connected component of the identity is $H_0
=\{\big(0,U(\varphi,0,\varphi)\big)\,|\,\varphi\in [0,2\pi]\}$ 
\cite{Makela03}. 
Now vortices and monopoles are possible and both of them are classified by 
integers. 
A general antiferromagnetic spinor can then be written as 
$\psi(\mathbf{r})=\sqrt{n(\mathbf{r})}\xi(\mathbf{r})$, where 
$\xi(\mathbf{r})=e^{i\theta(\mathbf{r})}U^{(1)}(\mathbf{r})\xi_{ref}$ 
and $U^{(1)}$ is shown in (\ref{U1}). Explicitly   
\begin{equation}\label{AF}
 \psi(\mathbf{r})=\sqrt{n(\mathbf{r})}e^{i\theta(\mathbf{r})} 
\left(\begin{array}{c}-e^{-i\alpha(\mathbf{r})}
\frac{1}{\sqrt{2}}\sin\beta(\mathbf{r})\\\cos\beta(\mathbf{r})\\e^{i\alpha(\mathbf{r})}
\frac{1}{\sqrt{2}}\sin\beta(\mathbf{r})\end{array}\right). 
\end{equation}

\subsubsection{Vortices}
As can be seen from the isotropy group $H$, 
vortices with winding numbers $m$ and $m+\frac{1}{2}$ are
possible ($m\in \mathbb{Z}$). Using an analysis similar to 
that of ferromagnetic condensates, a
vortex with a winding number $m$ can be shown to be given by (\ref{AF}) if 
$\theta$ fulfills the condition 
$\theta(r,z,0)=0,\theta(r,z,2\pi)= 2\pi m$. 
Other angles can be chosen freely as long as
$\xi(r,z,0)=\xi(r,z,2\pi)=|F=1,m_F=0\rangle$. 
For $m+\frac{1}{2}$ condensates the requirements are 
$\theta(r,z,0)=0,\theta(r,z,2\pi)= 2\pi (m+\frac{1}{2})$ and  
$\xi(r,z,0)=-\xi(r,z,2\pi)=|F=1,m_F=0\rangle$. 
The superfluid  
velocity is 
$\mathbf{v}=\frac{\hbar}{Mr}\frac{\partial\theta}{\partial
\varphi}\mathbf{e}_\varphi$ and it cannot vanish everywhere 
if there is a vortex in the system.  
The orbital angular momentum is $\mathbf{L}= N\hbar l\,
\mathbf{e}_z$, where 
$l$ is either $m$ or $m+\frac{1}{2}$.  
In the following we assume that   
$\psi(r,z,\varphi)=\sqrt{n(r,z)}\xi(\varphi)$.  
The spinor minimizing the energy of a vortex with winding number 
$m\in\mathbb{Z}$
 is obtained when 
$\alpha,\beta$ are constants and $\theta(\varphi)=m\varphi$. 
If we choose $\beta=0$ it becomes 
$\psi_m(r,z,\varphi)=\sqrt{n(r,z)} e^{im\varphi}|F=1,m_F=0\rangle$. 
A winding number $m+\frac{1}{2}$ vortex with minimum energy is given 
by (\ref{AF})
under the conditions that 
$\theta(r,z,2\pi)=(2m+\frac{1}{2})\pi,\alpha$ is constant and 
$\beta=\frac{\varphi}{2}$, 
\begin{equation}\label{AFvortex2}
\psi_{m+\frac{1}{2}}(r,z,\varphi)=\sqrt{n(r,z)}
 e^{i(m+\frac{1}{2})\varphi}
\left(\begin{array}{c}
-e^{-i\alpha}\frac{1}{\sqrt{2}}\sin\frac{\varphi}{2}\\
\cos\frac{\varphi}{2}\\
e^{i\alpha}\frac{1}{\sqrt{2}}\sin\frac{\varphi}{2}\end{array}\right).
\end{equation} 
Properties of vortices of the type (\ref{AFvortex2}) have been discussed 
before for example in \cite{Leonhardt00,Zhou01,Zhou03,Mueller04}.

\subsubsection{Monopoles} 

As explained in section 2, an expression for a winding number $m$ 
monopole can be found as an 
extension of a map $g_m:[0,2\pi]\rightarrow H_0$, where  
 $g_m$ determines the element of $\pi_1(H_0,\Id)$ 
with winding number $m$. Now we define    
$g_m(\varphi)=\left(0,U(\alpha_m(\varphi),0,\alpha_m(\varphi))\right)$, 
where $\alpha_m$ is such that $\alpha_m(0)=0,\alpha_m(2\pi)=2\pi m$. 
We define a map $\tilde{g}_m:D^2\rightarrow \mathbb{R}\times SU(2)$ 
such that $\tilde{g}_m(\pi,\varphi)=g_m(\varphi)$. 
Then 
$f_m:\mathbb{R}^3\setminus 
\{\mathbf{0}\}\rightarrow [\mathbb{R}\times SU(2)]/H$,   
$f_m(r,\theta,\varphi)=\tilde{g}_m(\theta,\varphi)H$ gives a monopole 
with winding number $m$. A general form for $\tilde{g}_m$ is 
$\tilde{g}_m(\theta,\varphi)=\big(\delta_m(\theta,\varphi),  
U(\alpha_m(\theta,\varphi),\beta_m(\theta,\varphi),
\alpha_m(\theta,\varphi))\big)$ with the conditions $\alpha_m(\pi,\varphi)
=\alpha_m(\varphi),\,\,  
\beta_m(\pi,\varphi)=\delta_m(\pi,\varphi)=0,$ which follow from 
$\tilde{g}_m(\theta=\pi,\varphi)=g_m(\varphi)$. From the continuity 
of $\tilde{g}_m$ it also follows that 
$\delta_m(0,\varphi)=0$ and 
$\beta_m(0,\varphi)=(2k+1)\pi$ with $k$ an arbitrary integer.

By a simple change of basis equation (\ref{AF}) can be cast in the form 
\begin{equation}\label{mono}
\psi_m(r,\varphi,\theta)=\sqrt{n(r,\theta,\varphi)}e^{i\delta}
\left(\begin{array}{c}
\cos\alpha_m\,\sin\beta_k\\
\sin\alpha_m\,\sin\beta_k\\
-\cos\beta_k
\end{array}\right).
\end{equation}
Here we have substituted $\delta,\alpha_m,\beta_k$ for
 $\theta,\alpha,\beta$, respectively. 
 If $\delta\equiv 0$ the above spinor is a real and normalized  
three-component vector. Then there is an integral 
equation 
\begin{equation}
w=\frac{1}{4\pi}\int_{0}^{2\pi}\int_0^{\pi}d\varphi\,d\theta\, \xi\cdot 
\left(\frac{\partial\xi}{\partial \theta}
\times \frac{\partial\xi}{\partial \varphi}\right)
\end{equation}
which gives the winding number $w$ of the monopole; see
 \cite{Rajaraman82} or \cite{Arafune75}. 
Using this equation one can confirm that by assuming $\delta_m\equiv 0$ 
and that $g_m$  
fulfills the conditions stated above,  
the winding number of the spinor  
in equation (\ref{mono}) is $m$. For a non-constant $\delta_m$ 
the winding number is the same, as it does not depend 
on the form of $\delta_m$. 
From equation (\ref{mono}) one sees that the superfluid velocity 
vanishes if and only if $\delta_m$ is a constant function. 
This follows from the fact that $\delta\equiv constant\iff \xi^\dag 
\nabla\xi\in \mathbb{R}^3$. Because  
$\mathbf{v}=-i\frac{\hbar}{M}\xi^\dag\nabla \xi\in\mathbb{R}^3$,  
it follows that $\mathbf{v}=\mathbf{0}$ if and only if $\delta_m$ 
is a constant.  

While it is instructive to know what are
 the general requirements for a spinor to represent a monopole, it 
is important to see that the typical expression for a monopole 
can also be obtained as a special case of above equations. If we choose 
$g_m$ such that  
$\delta_m\equiv 0,\alpha_m(\theta,\varphi)=m\varphi$ and 
 $\beta_m(\theta,\varphi)=(2k+1)(\pi-\theta)$, where $k$ is an arbitrary integer, 
all the requirements for $\delta_m,\alpha_m$ and $\beta_m$ are fulfilled and  
 the monopole becomes   
\begin{equation}
\psi_m(r,\varphi,\theta)=\sqrt{n(r,\theta,\varphi)}
\left(\begin{array}{c}
\cos(m\varphi)\,\sin[(2k+1)\theta]\\
\sin(m \varphi)\,\sin[(2k+1)\theta]\\
\cos[(2k+1)\theta]
\end{array}\right).
\end{equation}
Choosing $k=0$ gives the usual expression for a monopole. 
It has been presented in the context of BEC before for example 
in \cite{Stoof01,Ruostekoski03}, but the requirements for a general 
expression of a monopole and its properties have not been discussed before.   
A monopole with $m=1,k=0$ 
has been numerically studied in \cite{Stoof01,Ruostekoski03}.

\subsection{Cyclic states}
The defects of the $C0$ phase have already been discussed 
in \cite{Makela03}, so here  we concentrate on the defects of the   
$C1$ phase. The isotropy group $H$ has not been calculated before, so
we have to do it here. 
For the $SU(2)$-matrices we now use the $U$-representation 
shown in (\ref{U}). The reference spinor is chosen to be 
$C1=\frac{1}{\sqrt{3}}(1,0,0,\sqrt{2},0)^T$ and 
a general expression for  a spinor is 
\begin{equation}
\fl
\psi(r,z,\varphi)=\sqrt{n(r,z,\varphi)}e^{i\theta}\left(
\begin{array}{c}
\frac{e^{-2i(\alpha+\gamma)}}{\sqrt{3}}[ \cos^4\frac{\beta
    }{2}-\sqrt{2}e^{3i\gamma}
\sin^2\frac{\beta}{2}\sin\beta]\\
 \frac{e^{-i(\alpha+2\gamma)}}{\sqrt{6}}[e^{3i\gamma}(\cos\beta-\cos
    (2\beta))+\sqrt{2}\cos^2\frac{\beta}{2}\sin\beta]\\
 \frac{e^{-2i\gamma}}{4}\sin\beta\,[-4e^{3i\gamma}\cos\beta 
+ \sqrt{2}\sin\beta] \\
 \frac{e^{i(\alpha-2\gamma)}}{\sqrt{6}}[e^{3i\gamma}(\cos\beta+\cos(2\beta))
+\sqrt{2}\sin^2\frac{\beta}{2}\sin\beta]\\ 
\frac{e^{2i(\alpha-\gamma)}}{\sqrt{3}}[ \sin^4\frac{\beta}{2}
+\sqrt{2}e^{3i\gamma}
\cos^2\frac{\beta}{2}\sin\beta]
\end{array}\right).
\end{equation} 
Equating this with the reference spinor 
$C1$ shows that the isotropy group 
is $H=\{\left(\frac{2\pi}{3}(2n+3m),U(\frac{2\pi
  n}{3},0,0)\right)\,|\, m\in\mathbb{Z}, \, n=0,\ldots 5\}
\subset \mathbb{R}\times SU(2)$ and $H_0=(0,\mathbb{I})$. 
This means that only vortices,   
not monopoles, are topologically stable. 
$H/H_0=H$ is isomorphic to the group $\mathbb{Z}\times \mathbb{Z}_6$, 
the isomorphism is given by the map $\left(\frac{2\pi}{3}(2n+3m),U(\frac{2\pi
n}{3},0,0)\right)\mapsto (m,n)\in \mathbb{Z}\times \mathbb{Z}_6 $. 
Therefore vortices are classified by two winding numbers $(m,n)$ and those with 
different $m$ cannot be converted into one another. However, for  
vortices with winding numbers 
$(m,n)$ and $(m,n+6)$ this is possible. 

A vortex with winding numbers $(m,n)$ is given by
$f_{m,n}:\mathbb{R}^3\setminus\mathbb{R}\mathbf{e}_z\rightarrow
\mathbb{R}\times SU(2)$ such that $f_{m,n}(r,z,0)=\Id$ and 
$f_{m,n}(r,z,2\pi)=\left(\frac{2\pi}{3}(2n+3m),U(\frac{2\pi
  n}{3},0,0)\right)$ for all $z\in\mathbb{R},r>0$. Now we define 
$f_{m,n}$ by   
$f_{m,n}(r,z,\varphi)=\left(\frac{\varphi}{3}(2n+3m),U(\frac{\varphi
  n}{3},0,0)\right)$.  
The corresponding order-parameter is 
\begin{equation}
\psi(r,z,\varphi)=\sqrt{n(r,z)}\left(
\begin{array}{c}
\frac{1}{\sqrt{3}}e^{i\varphi m}\\0\\0\\\sqrt{\frac{2}{3}}
 e^{i\varphi (m+n)}\\0
\end{array}\right).
\end{equation}
This is also the order-parameter that minimizes the vortex energy.  
The superfluid velocity and angular momentum are              
$\mathbf{v}=-\frac{\hbar}{Mr}(m+\frac{2n}{3})\,\mathbf{e}_\varphi$ 
and $\mathbf{L}=N\hbar(m+\frac{2n}{3})\mathbf{e}_z$.

\section{Effects of external magnetic field and magnetization}
So far we have assumed that there is no external magnetic field present. 
Now we consider the situation where the condensate is placed in a magnetic
 field directed parallel to the $z$-axis.
To the second order in the strength $B$ of the magnetic field the energy 
from the field can be written as \cite{Pethick02}
\begin{equation}\label{EB}
E_B[\psi]=\int d^3 r\, 
n(\mathbf{r})\big[\gamma B (\mathbf{r})
\langle F_z\rangle +\epsilon B^2(\mathbf{r})
\langle F_z^2\rangle \big],
\end{equation} 
where $\gamma$ and $\epsilon$ are constants. 
For ${}^{87}$Rb and ${}^{23}$Na these are $\gamma=\pm \frac{\mu_B}{2}$
 and $\epsilon=\mp\frac{\mu_B^2}{4\Delta E_{hf}}$. Here 
$\mu_B$ is the Bohr magneton, $\Delta E_{hf}$ is the hyperfine
 splitting between $F=2$ and $F=1$ states and upper (lower) 
sign refers to $F=2$ ($F=1$).   
In the presence of a magnetic field the conservation 
of magnetization has to be taken into attention. If the magnetic 
field is parallel to the $z$-axis the magnetization is defined by 
$M=\int d^3 r\, n(\mathbf{r})\langle F_z\rangle
=\int d^3 r\, n(\mathbf{r})\sum_{j=-F}^F j\,|\xi_j
 (\mathbf{r})|^2$ $(\hbar=1)$. Magnetization can have any value between 
$-N F$ and $N F$, where $N$  is the particle number.  
In the absence of an external magnetic field the magnetization is 
a meaningless quantity because of the lack of a well-defined quantization 
axis of the hyperfine spin. 
In the presence of a magnetic field 
the magnetization is a well-defined and often conserved quantity since the 
 collisions that do not conserve magnetization occur usually  
in a timescale much 
 longer than a typical lifetime of the condensate. The conservation 
of magnetization has been
experimentally verified \cite{Schmaljohann04,Chang04}. 
Thus the energy of the system
should be minimized under the assumption of fixed $M$. 
If we consider  
weak values of the magnetic field it is enough to include 
only the first term in equation (\ref{EB}) 
in the energy. If the magnetic field is spatially
constant this term is just  magnetization multiplied by a constant.     
Thus the ground state is determined by the value of 
magnetization, and does not depend on the strength of the 
magnetic field, as long as the quadratic term in the 
magnetic field is negligible. 

If the external field is absent  the minimum 
of energy is  obtained (if $F=1$) 
when either $|\langle \mathbf{F}\rangle|=0$ or 
$|\langle \mathbf{F}\rangle|=1$. If external field is present  
the energy has to be minimized under the assumption 
of conserved magnetization. For a ferromagnetic $F=1$ condensate 
any magnetization can be produced while 
$|\langle \mathbf{F}\rangle|=1$. Thus in this case 
the ground state is obtained if one finds an order-parameter 
which produces the given magnetization and fulfills 
the condition  $|\langle \mathbf{F}\rangle|=1$.
On the other hand, if $|\langle \mathbf{F}\rangle|=0$ the 
 only possible value for magnetization is zero.    
 If the magnetization is non-zero and the system is 
antiferromagnetic, the
 condition  $|\langle \mathbf{F}\rangle|=0$ cannot be satisfied 
everywhere. In this case the calculation of the ground state 
is more difficult. The situation is similar in the cyclic and polar
(antiferromagnetic) phases of $F=2$ system.

Next we assume that only the term linear in the magnetic field is included 
in the energy and show that the previously calculated 
vortex of a ferromagnetic $F=1$ condensate is possible for an arbitrary
value of the magnetization. 
Direct calculation shows that for the vortex of equation
(\ref{F1vortex}) the magnetization is $N\cos^2\beta$. Any magnetization
between $0$ and $N$ can be obtained as 
$\beta$ is varied between $\frac{\pi}{2}$ and $0$. The vortex 
with minimum energy is obtained when magnetization vanishes. 
Because equation (\ref{F1vortex}) is obtained using $|F=1,m_F=1\rangle$ 
as the reference order-parameter, it is not surprising that 
positive values of magnetization are favoured by it. A vortex 
configuration which has negative magnetization can be obtained 
by choosing  $|F=1,m_F=-1\rangle$ as the reference 
spinor and calculating a representative of a vortex as before.
The magnetization 
turns out to be equal to $-N\cos^2\beta$. 
The vortices corresponding to different choices 
of the reference order-parameter have different magnetizations, 
but if the magnetization is allowed to vary, these vortices   
can be converted continuously into one another. 
Thus they are similar from topological point of view.

Next we see that setting magnetization fixed makes new kind 
of defects stable.  
If we choose $\beta=\frac{\pi}{2}$ in equation (\ref{F1}), 
multiply the spinor  
by $e^{i\theta}$ and redefine $\alpha\rightarrow \alpha+\frac{\pi}{2}$, 
we get 
\begin{equation}\label{F1prime}
\psi(r,z,\varphi)=\sqrt{n(r,z,\varphi)}\left(
\begin{array}{cc}
&e^{i\theta}\cos^2\frac{\tau}{2} \\
 &\frac{1}{\sqrt{2}}e^{i(\alpha+\theta)}\sin\tau\, 
\\ &e^{i(2\alpha+\theta)}\,\sin^2 \frac{\tau}{2}  
\end{array}\right).
\end{equation} 
This is an alternative parametrization for a ferromagnetic spinor. 
From the form 
 of the order-parameter one might assume that a vortex is obtained 
 if $\theta=m\varphi$ and $\alpha=n\varphi$ with $m,n$ integers. 
Like before, we assume that the vortices are straight and located on the $z$-axis. 
We also require that $n=n(r,z),\tau=\tau(r,z)$. 
The latter assumptions mean that the particle density 
of each spin component is cylindrically symmetric\footnote{These 
kinds of vortices are called {\it axisymmetric} in \cite{Isoshima02,Mizushima02b}}.
The magnetization is $M=\int d^3\mathbf{r} 
\,n(\mathbf{r})\cos\tau(r,z)$, which shows that 
 (\ref{F1prime}) can produce any magnetization if $\tau$ 
is chosen properly.    
Topologically the decay of a vortex in one component can now only be achieved 
by converting all atoms from that spin state into other states.  
From the above spinor one sees that always two components vanish
simultaneously. When this happens magnetization is $\pm N$. 
 Thus if the magnetization is fixed and different from 
$\pm N$, one cannot make any spin component vanish while keeping 
the magnetization fixed.   
If the magnetization is different from $\pm N$, the order-parameter space 
is $S^1\times S^1$, since the phase of two components can 
be chosen freely. The dynamical stability and dynamics of these 
type of vortices has been studied in \cite{Isoshima01,Isoshima02,Mizushima02a,
Mizushima02b,Martikainen02}. However, their topological 
stability following from 
the conservation of magnetization has not been pointed out before.  
 If the order-parameter is of the form (\ref{F1prime}), vortices 
appear in each spin component separately. This means that the total density 
does not have to vanish, unless there is a vortex in every component. 
If there is one spin component without vortex the system is a coreless vortex. 
For example by choosing $m=0,n=1$ or $m=1,n=-1$ one gets an expression for a 
coreless vortex. 

If the magnetization is not a conserved quantity vortices with even 
$m$ can be converted into a uniform configuration and 
those with odd $m$ are equivalent with the winding number one vortex
 of equation (\ref{F1vortex}). This is because in zero field $\tau$
 can be converted into a map for which $\tau(\mathbf{r})=0$ holds for
 all $\mathbf{r}$. Then the spinor becomes 
$e^{im\varphi}|F=1,m_F=1\rangle$. This in turn is equivalent 
with a vortex with winding number zero or one for $m$ even or odd, 
respectively; see \cite{Ho98}. 
 
Above we have assumed that the particle densities of different spin components 
are symmetric. 
It is possible that a deviation from this allows a continuous decay 
of vortex. This is  also indicated by numerical studies \cite{Mizushima02b}.  
 
If the magnetic field is strong,  
the term quadratic in the magnetic field has to be included in the
energy. Also in this case the magnetization is conserved, 
and vortices which are possible when only the linear term is 
included remain topologically stable.

\section{Creation of defects}
Next we propose a method to create 
vortices in a ferromagnetic condensate. These vortices are topologiccally 
stable also in the absence of an external magnetic field.  
In \cite{Ogawa02,Mottonen02} a way to create   
a vortex exploiting the spin degree of freedom has been studied. 
In this method a condensate in a low-field 
seeking state is prepared in an Ioffe-Pritchard trap.  
Initially the magnetic field in the $z$-direction is assumed 
to be much stronger than the magnetic field $B_\perp$ in the $xy$-plane. 
The $z$-component of the field $B_z$ is then reversed slowly, 
while keeping $B_\perp$ fixed. In this way 
a vortex with winding number $2F$ is created,  
$F$ being the hyperfine spin of the condensate atoms. 
The feasibility of this method has been experimentally 
verified using a ${}^{23}$Na condensate  
prepared in the low-field seeking states
$|F=1, m_F=-1\rangle$ 
or $|F=2, m_F=+2\rangle$ \cite{Leanhardt02}.  
Since a vortex with a winding number larger than one 
is energetically unstable against decay into 
winding number one vortices, it is presumed that 
vortices created this way will split. 
In the case of $F=1$ ${}^{23}$Na condensate 
this has been seen to occur \cite{Shin04}, and 
it is expected to happen also in the $F=2$ case.  
In addition to creating defects which are stable in the presence 
of magnetic field, we propose how a modification of this
 method can be used to 
create $2F$ vortices with winding number 
one in a ferromagnetic condensate with hyperfine spin $F$.  
These vortices are stable in the absence of an external magnetic 
field. To create these vortices, in addition to an Ioffe-Prichard trap,  
an optical trap 
is needed.  
First a vortex with winding number $2F$ 
is created in the previously described way.  
Then   
one waits until the vortex decays 
into winding number one vortices. 
Then $B_\perp$ is turned off, and the optical trap is turned on 
simultaneously. This   
does not change the spin state because it is assumed that $B_z\gg B_\perp$. 
The remaining 
field in the $z$-direction can then be reduced 
to zero, and vortices are allowed to 
evolve in the optical trap. 
If the magnetic trap is turned 
off before the vortex has split, the vortex can in principle continuously 
convert into a non-vortex state. This is possible,  
for example, if the vortex reverses the rotation 
which was used to create it. This is prevented by letting the vortex 
split before turning off the Ioffe-Pritchard trap.    
In experiments there are usually stray ac magnetic fields, so    
the creation of zero-field defects requires effective 
magnetic shielding, which can be troublesome.
Achieving this would, however, be rewarding, as 
it would allow one to change the topological stability  
of vortices as a function of the magnetic field.

The creation of monopoles is not as straightforward 
as that of vortices, but a method for this has been 
proposed in  \cite{Chang02}.

\section{Conclusions}
In this paper we have first derived a systematic
 way to create explicit expressions 
for vortices and monopoles. 
This method requires the calculation of the first 
and second homotopy groups of the order-parameter space $G/H$.  
This can be achieved by using equations 
obtained from the exact sequence of relative homotopy groups. 
After this the expressions for defects can be constructed 
by finding suitable mappings from the physical space 
into the group $G$.

 We have created 
examples of vortices and monopoles 
in spinor Bose-Einstein
condensates using this method. Especially the defects in zero
external magnetic field have been discussed. We have presented  
examples of vortices 
in ferromagnetic $F=1$ and $F=2$ condensates and    
 vortices and monopoles in antiferromagnetic 
$F=1$ condensate. We also pointed out that the order-parameter 
space of the cyclic phase of $F=2$ condensate consists of two disconnected sets. 
The properties of one of the sets have been studied previously in \cite{Makela03}. 
Here we calculated the topological defects of the other set and showed that vortices classified by 
$\mathbb{Z}\times\mathbb{Z}_6$ are topologically stable, whereas monopoles are not possible. 
Also the superfluid velocity induced by the defects is examined. 
It has been shown that in a ferromagnetic condensate with hyperfine spin $F$  
the presence of a vortex with winding number 
$F$ does not have to induce non-zero superfluid velocity or orbital angular momentum.

We have also studied the effect of a magnetic field, concentrating on
a ferromagnetic $F=1$ condensate. It has been shown 
that a vortex which is topologically stable
 in the absence of a magnetic field is also 
possible if the magnetization has a fixed value.  
This means that it is also possible 
if there is a homogeneous magnetic field present. 
In addition to this, we have found out that the conservation of  
magnetization may stabilize   
vortices which are not topologically  stable 
if the magnetization can vary freely. 
Thus there can be a transition 
from a vortex state to a non-vortex state as the magnetic field 
strength is lowered to zero. 
Finally a method to create vortices which are topologically 
stable in the absence of a magnetic field has been suggested.  

\setcounter{section}{1}
\ack
Part of this work was done during a visit to NORDITA. I thank 
Chris Pethick for his hospitality. 
The author is grateful to 
K.-A. Suominen for careful reading of the manuscript 
and useful suggestions. Discussions with Yunbo Zhang are 
also appreciated. 
 This work was supported by the Academy of Finland (project 206108) and 
the Vilho, Yrj\"o and Kalle V\"ais\"al\"a  
Foundation (Finnish Academy of Science and Letters).

\appendix
\section{Spin matrices and rotation operators}
The spin matrices for $F=1$ are 
\begin{equation}
\fl 
F_x=\frac{1}{\sqrt{2}}\left(\begin{array}{cccc} 
0&1&0\\1&0&1\\0&1&0\end{array}\right),\quad
F_y=\frac{1}{\sqrt{2}}\left(\begin{array}{cccc} 
0&-i&0\\i&0&-i\\0&i&0\end{array}\right),\quad
F_z=\left(\begin{array}{cccc} 
1&0&0\\0&0&0\\0&0&-1\end{array}\right),  
\end{equation}
and those of $F=2$ are 
\begin{equation}
\fl
\begin{array}{ll}
&F_x=\frac{1}{2}\left(\begin{array}{ccccc} 
0&2&0&0&0\\2&0&\sqrt{6}&0&0\\0&\sqrt{6}&0&\sqrt{6}&0\\
0&0&\sqrt{6}&0&2\\0&0&0&2&0
\end{array}\right),\quad
F_y=\frac{i}{2}\left(\begin{array}{ccccc} 
0&-2&0&0&0\\2&0&-\sqrt{6}&0&0\\0&\sqrt{6}&0&-\sqrt{6}&0\\
0&0&\sqrt{6}&0&-2\\0&0&0&2&0
\end{array}\right),\\
&F_z=\left(\begin{array}{ccccc} 
2&0&0&0&0\\0&1&0&0&0\\0&0&0&0&0\\0&0&0&-1&0\\0&0&0&0&-2
\end{array}\right). 
\end{array}
\end{equation}
Here we have set $\hbar =1$. Depending on the 
system studied, the  elements of $SU(2)$ have been written either in the form 
  
\begin{equation}
\label{U}
\fl
U(\alpha,\beta,\gamma)=e^{-i\alpha F_z}e^{-i\beta F_y}
e^{-i\gamma F_z}=
\left(\begin{array}{cc}\cos\frac{\beta}{2}e^{-i(\alpha+\gamma)/2}
  &-\sin\frac{\beta}{2}e^{i(\gamma-\alpha)/2}
\\\sin\frac{\beta}{2}e^{-i(\gamma-\alpha)/2} &
\cos\frac{\beta}{2}e^{i(\alpha+\gamma)/2} \end{array}\right), 
\end{equation}
or in the form
\begin{equation}
\label{V}
\fl
V(\tau,\alpha,\beta)=e^{-i\frac{\tau}{2}\mathbf{n}\cdot\mathbf{\sigma}}=
\left(\begin{array}{ccc}
\cos\frac{\tau}{2}-i\sin\frac{\tau}{2}\cos\beta &
-ie^{-i\alpha}\sin\frac{\tau}{2}\sin\beta\\
-ie^{i\alpha}\sin\frac{\tau}{2}\sin\beta&
\cos\frac{\tau}{2}+i\sin\frac{\tau}{2}\cos\beta
\end{array}\right),\end{equation}
where $\mathbf{n}=(\cos\alpha\sin\beta,\sin\alpha\sin\beta,\cos\beta)$ 
and $\sigma=(\sigma_x,\sigma_y,\sigma_z)$ is a vector formed from 
Pauli matrices. The $V$-matrix has been used when discussing the ferromagnetic 
condensates, otherwise the $U$-matrix has been used.   
$2F+1$ -dimensional irreducible representations of these matrices are given 
by the maps $U(\alpha,\beta,\gamma)\mapsto
U^{(F)}(\alpha,\beta,\gamma)$ and  
$V(\tau,\alpha,\beta)\mapsto V^{(F)}(\tau,\alpha,\beta)$. 
Here $U^{(F)}(\alpha,\beta,\gamma)=\exp(-i\alpha F_z) \exp(-i\beta F_y)
\exp(-i\gamma F_z), V^{(F)}(\tau,\alpha,\beta)=\exp(-i\tau\,
\mathbf{n}\cdot\mathbf{F})$
 and $\mathbf{F}$ is the spin operator of spin $F$ system.  
In this paper we need explicit expressions for $U^{(F)}$ and
$V^{(F)}$ for $F=1,2$.  
For $F=1$ these are 

\begin{equation}
\label{U1}
\fl
U^{(1)}(\alpha,\beta,\gamma)=\left(\begin{array}{cccc}
&e^{-i(\alpha+\gamma)}\cos^2\frac{\beta}{2}
 &-e^{-i\alpha}\frac{1}{\sqrt{2}}\sin\beta
 &e^{-i(\alpha-\gamma)}\sin^2\frac{\beta}{2}\\
&e^{-i\gamma}\frac{1}{\sqrt{2}}\sin\beta&\cos\beta & -e^{i\gamma}
\frac{1}{\sqrt{2}}\sin\beta\\
 &e^{i(\alpha-\gamma)}\sin^2\frac{\beta}{2}&
 e^{i\alpha}\frac{1}{\sqrt{2}} \sin\beta
& e^{i(\alpha+\gamma)}\cos^2\frac{\beta}{2}
\end{array}\right)
 \end{equation}

and 
\begin{equation}
\label{V1}
\begin{array}{ll}
\fl 
V^{(1)}(\tau,\alpha,\beta)=\\
\fl\left(
\begin{array}{cccc}
&\left(\cos\frac{\tau}{2} - i\cos\beta\,\sin\frac{\tau}{2}\right)^2 
& e^{-i \alpha }\sin \beta 
 \frac{\left(-1+\cos \tau\right)\cos\beta-i\sin \tau}{\sqrt{2}}
  \\
 &-\sqrt{2}e^{i\alpha}\sin\beta\,\sin\frac{\tau}{2}\,
 \left(i\cos\frac{\tau}{2}+\cos\beta\,\sin \frac{\tau}{2}\right) 
&\cos^2\frac{\tau}{2}+ \cos
  (2\beta)\,\sin^2\frac{\tau}{2} 
\\
 &-e^{2i\alpha}\,\sin^2 \frac{\tau}{2}\sin^2\beta  &
 -e^{i\alpha }\sin \beta \frac{\left(-1 + \cos \tau \right) 
\cos \beta +i\sin \tau}{\sqrt{2}}
\end{array}\right.\\ 
\\
\fl
\left.
\begin{array}{ccc}
& -e^{-i 2\alpha }\sin^2\frac{\tau}{2}\sin^2\beta \\
& -e^{-i\alpha}\sin\beta
 \frac{\left(-1 + \cos \tau\right) \,\cos\beta  + i\sin \tau}{\sqrt{2}} \\
& \left(\cos \frac{\tau}{2} + i\cos\beta \,\sin \frac{\tau}{2}\right)^2 \\
\end{array}
\right).
\end{array}
\end{equation}

For $F=2$ the spin rotation matrix is  
\begin{equation}
\begin{array}{lll}
\fl U^{(2)}(\alpha,\beta,\gamma) =\\
\fl
\left(\begin{array}{ccccc}
e^{-2i(\alpha+\gamma)}\cos^4\frac{\beta}{2}&-e^{-i(2\alpha+\gamma)}
\sin\beta\cos^2\frac{\beta}{2}& e^{-2i\alpha}\frac{\sqrt{6}}{4}
\sin^2\beta
\\
e^{-i(\alpha+2\gamma)}\sin\beta\cos^2\frac{\beta}{2}&e^{-i(\alpha+\gamma)}
\frac{1}{2}
(\cos\beta +\cos
2\beta)& -e^{-i\alpha}\frac{\sqrt{6}}{4}\sin 2\beta
 \\e^{-2i\gamma} \frac{\sqrt{6}}{4}
\sin^2\beta&e^{-i\gamma}
 \frac{\sqrt{6}}{4}\sin 2\beta &\frac{1}{4}(1+3\cos 2\beta)
\\e^{i(\alpha-2\gamma)}
 \sin\beta\sin^2\frac{\beta}{2}
 & e^{i(\alpha-\gamma)}\frac{1}{2}(\cos \beta -\cos 2\beta)&e^{i\alpha}
\frac{\sqrt{6}}{4}\sin 2\beta
&
\\e^{2i(\alpha-\gamma)}\sin^4\frac{\beta}{2}&e^{i(2\alpha-\gamma)}
\sin\beta\sin^2\frac{\beta}{2} &e^{2i\alpha}\frac{\sqrt{6}}{4}
\sin^2\beta
\end{array}\right.\\
\\
\fl
\left. \begin{array}{ccc}
&-e^{-i(2\alpha-\gamma)}\sin\beta\sin^2\frac{\beta}{2}&e^{-2i(\alpha-\gamma)}
\sin^4\frac{\beta}{2}\\
&
e^{-i(\alpha-\gamma)}\frac{1}{2}(\cos\beta-\cos
2\beta)&-e^{-i(\alpha-2\gamma)}
\sin\beta\sin^2\frac{\beta}{2}\\
&-e^{i\gamma}\frac{\sqrt{6}}{4}\sin 2\beta
 &e^{2i\gamma}\frac{\sqrt{6}}{4}\sin^2\beta\\ 
 &e^{i(\alpha+\gamma)}\frac{1}{2} (\cos\beta +\cos
2\beta)&-e^{i(\alpha+2\gamma)}\sin\beta\cos^2\frac{\beta}{2}
\\
&e^{i(2\alpha+\gamma)}
\sin\beta\cos^2\frac{\beta}{2}
&e^{2i(\alpha+\gamma)} \cos^4\frac{\beta}{2}
\end{array} \right). 
\end{array}
\end{equation}

\end{document}